\documentclass{aa}
\usepackage[varg]{txfonts}
\usepackage{natbib}

\newcommand{\vect}[1]{\boldsymbol{#1}}
\renewcommand{\thesubsection}{\arabic{subsection}}
\usepackage{makecell}
\usepackage{hyperref}
\usepackage{verbatim}
\usepackage{pbox}
\usepackage{physics}

\newcommand{\mpe}{Max Planck Institute for extraterrestrial Physics, Giessenbachstr., 85748 Garching, Germany, \email{idelw@mpe.mpg.de}}
\newcommand{\ipag}{Univ. Grenoble Alpes, CNRS, IPAG, F-38000 Grenoble, France}

\begin{document}

\title{Super-Keplerian Equatorial Outflows in SS 433 \thanks{Based on observations collected at the European Southern Observatory, Chile, Program ID 099.D-0666(A).}}
\subtitle{Centrifugal Ejection of the Circumbinary Disk} 

\author{Idel Waisberg\inst{1} \and Jason Dexter\inst{1} \and Pierre Olivier-Petrucci\inst{2} \and Guillaume Dubus\inst{2} \and Karine Perraut\inst{2}} 
\institute{\mpe \and \ipag} 

\abstract {The microquasar SS 433 is the only known steady supercritical accretor in the Galaxy. It is well-known for its relativistic baryonic jets, but the system also drives equatorial outflows. These have been routinely detected in radio images, and components associated with a circumbinary disk have also been suggested in optical emission lines.} 
{We wish to spatially resolve the regions producing the stationary emission lines of SS 433 to shed light on its circumbinary structure and outflows. With an estimated binary orbit size $\lesssim 0.1 \text{ mas}$, this requires optical interferometry.}  
{We use the optical interferometer VLTI+GRAVITY to spatially resolve SS 433 in the near-infrared K band at high spectral resolution ($R\approx 4000$) on three nights in July 2017. This is the second such observation, after the first one in July 2016.}
{The stationary Br$\gamma$ line in the 2017 observation is clearly dominated by an extended $\sim 1 \text{ mas} \sim 5 \text{ AU}$ circumbinary structure perpendicular to the jets and with a strong rotation component. The rotation direction is retrograde relative to the jet precession, in accordance with the slaved disk precession model. The structure has a very high specific angular momentum and is too extended to be a stable circumbinary disk in Keplerian rotation; interpreting it as such leads to a very high enclosed mass $M \gtrsim 400 M_{\odot}$. We instead interpret it as the centrifugal ejection of the circumbinary disk, with the implication that there must be an efficient transfer of specific angular momentum from the binary to the disk. We suggest that the equatorial outflows sometimes seen in radio images result from similar episodes of circumbinary disk centrifugal ejection. In addition to the equatorial structure, we find a very extended $\sim 6 \text{ mas} \sim 30 \text{ AU}$ spherical wind component to the Br$\gamma$ line: the entire binary is engulfed in an optically thin spherical line emission envelope.}
{} 


\maketitle

\section{INTRODUCTION}
\label{introduction}

The extreme emission-line object SS 433 \citep{Stephenson77,Clark78} was the first microquasar discovered, from its broad, red/blueshifted hydrogen and helium emission lines moving across its optical spectrum \citep{Margon79} and produced by relativistic, precessing baryonic jets moving at $0.26 c$ \citep{Fabian79,Margon84}. The jets are also seen in emission lines of highly ionized metals in X-rays \citep[e.g.][]{Marshall13} and as moving knots \citep[e.g.][]{Vermeulen93} and large-scale corkscrew structure in radio \citep[e.g.][]{Blundell04}. SS 433 is the only known Galactic manifestation of a \textit{steady} super-Eddington accretion disk, which outshines its donor star at all wavelengths and drives powerful outflows, manifested not only in the jets but also in strong, broad and complex "stationary" (in wavelength) emission lines. The estimated mass outflow $\dot{M} \sim 10^{-4} M_{\odot}\text{/yr}$ \citep{Shklovskii81,Fuchs06} establishes SS 433 as an outflow-regulated supercritically accreting system ($\frac{\dot{M}}{\dot{M_{Eddington}}} \sim 500$ for a $10 M_{\odot}$ black hole). For a review of SS 433's fascinating properties, see \cite{Fabrika04}. 

Although famous for its jets, one of the more exotic aspects of SS 433 are its equatorial outflows. The presence of an equatorial excretion flow from the accretion disk was proposed to explain the photometric and eclipsing behavior of SS 433 by \cite{Zwitter91}, possibly fed from the Lagrangian point behind the compact object \citep{Fabrika93}. The equatorial outflows were later detected in high-resolution radio images as outflowing emission knots at anomalous position angles, close to perpendicular to the jets \citep{Paragi99}. \cite{Blundell01} later detected a smooth, extended ($\sim 40$ mas) equatorial structure in radio images, calling it the ``radio ruff". A collection of further observations \citep{Paragi02,Mioduszewski04} suggested that the orientation of the outflows is roughly perpendicular to but span a larger angle range ($\sim 70^\circ$) than the precessing jets ($\approx 40^\circ$) \citep{Doolin09}. 

On the other hand, the presence of equatorial, circumbinary material has also been inferred from the double-peaked shapes that often appear in the optical emission lines. \cite{Fillipenko88} ascribed the double peaks with half-separation $\sim 150$ km/s in the high-order Paschen lines to an accretion disk (deriving a rather low mass, suggestive of a neutron star, for the compact object), but also recognized that the structure may instead arise in a circumbinary disk that, if in Keplerian rotation, would imply a much larger $\gtrsim 40 M_{\odot}$ enclosed mass. \cite{Robinson17} presents a similar analysis of the higher-order Brackett lines, assigning them to an accretion disk and favoring a neutron star as the compact object. On the other hand, based on decomposing the line profiles with several different Gaussian components, \cite{Blundell08} concluded that the H$\alpha$ emission line arises from a combination of a disk wind and a circumbinary disk, manifest as stable Gaussian components with half-separation $\gtrsim 200$ km/s. This velocity was interpreted as evidence that the total system mass must be large ($\gtrsim 40 M_{\odot}$) and the compact object must be a massive $\gtrsim 16 M_{\odot}$ black hole. Gaussian components arising from a circumbinary disk have also been suggested in the Br$\gamma$ line \citep{Perez09} and again in H$\alpha$ and He I \citep{Bowler10}. \cite{Cherepashchuk18} argues that the double-peaked structure must indeed arise from extended material because the wings of the line are not eclipsed (as would be expected for an accretion disk; SS 433 is an eclipsing binary). On the other hand, radial velocity measurements, notably extremely challenging in SS 433 due to the complexity of the emission lines and lack of clear stellar signatures, tend to favor lower masses $2 - 5 M_{\odot}$ for the compact object \citep{Hillwig08,Kubota10}. The relation between the circumbinary structure detected in optical emission lines and the equatorial outflows seen in radio is not clear. It has been suggested that the former might feed the latter \citep{Blundell08,Doolin09}. 

With an orbital period $P_{orb} = 13.1$ days \citep{Goranskii98} and distance $d = 5.5 \text{ kpc}$ \citep{Blundell04}, the semi-major axis of SS 433 is $a_{orb} = \left(\frac{M}{40 M_{\odot}}\right)^{1/3} \times 0.07 \text{ mas}$, where $M$ is the total binary mass. Spatially resolving the optical emission requires sub-mas resolution. This is beyond the capabilities of even future extremely large telescopes, but is achievable through spectro-differential optical interferometry. 
In \cite{GRAVITYSS43317} (Paper I) we presented the first such observations taken during commissioning of the GRAVITY instrument \citep{GRAVITY17} in July 2016 at the Very Large Telescope Interferometer (VLTI), which works in the near-infrared K band. These observations spatially resolved the near-infrared continuum as well as the stationary double-peaked Br$\gamma$ line. The interferometric signature across the latter showed a complex structure dominated by emission in the jet direction, suggestive of a bipolar outflow.

Here we report on a second set of observations of SS 433 with GRAVITY in July 2017, which clearly revealed equatorial emission with a strong rotation component. In Section 2, we summarize the observations and data reduction. The analysis of the K-band near-infrared continuum is presented in Section 3, whereas Section 4 describes the results on the stationary Br$\gamma$ line. Finally, Section 5 presents the conclusions. 

We often quote the results in mas since that is the actually measured unit. For convenience, we quote $1 \text{ mas} \leftrightarrow 8.23 \times 10^{13} \text{ cm} = 1180 R_{\odot} = 5.5 \text{ AU}$, assuming a distance $d = 5.5 (\pm 0.2)$ kpc derived from radio images using the aberration induced by light traveltime effect between the two jets \citep{Blundell04}. The GAIA DR2 distance $4.6 \pm 1.3 \text{ kpc}$ \citep{Luri18} is consistent with this value. 

\section{OBSERVATIONS AND DATA REDUCTION} 
\label{Observations and Data Reduction}

SS 433 ($K\approx8$) was observed with GRAVITY \citep{GRAVITY17} with the Unit Telescopes (UT) on VLTI on three nights over a period of four days in July 2017. Half of the K band light of SS 433 itself was directed to the fringe tracker (FT), which operates at $> 1000$ Hz to stabilize the fringes in the science channel (SC), allowing coherent integration over detector integration times of 10s in high spectral resolution ($R \approx 4000$). The FT operates in low resolution ($R\approx 20$) with five channels over the K band. The data were obtained in split polarization mode. The adaptive optics (AO) was performed at visual wavelength using SS 433 itself as the AO guide star ($V \approx 14$). 

\begin{figure}[]
\centering
\includegraphics[width=\columnwidth]{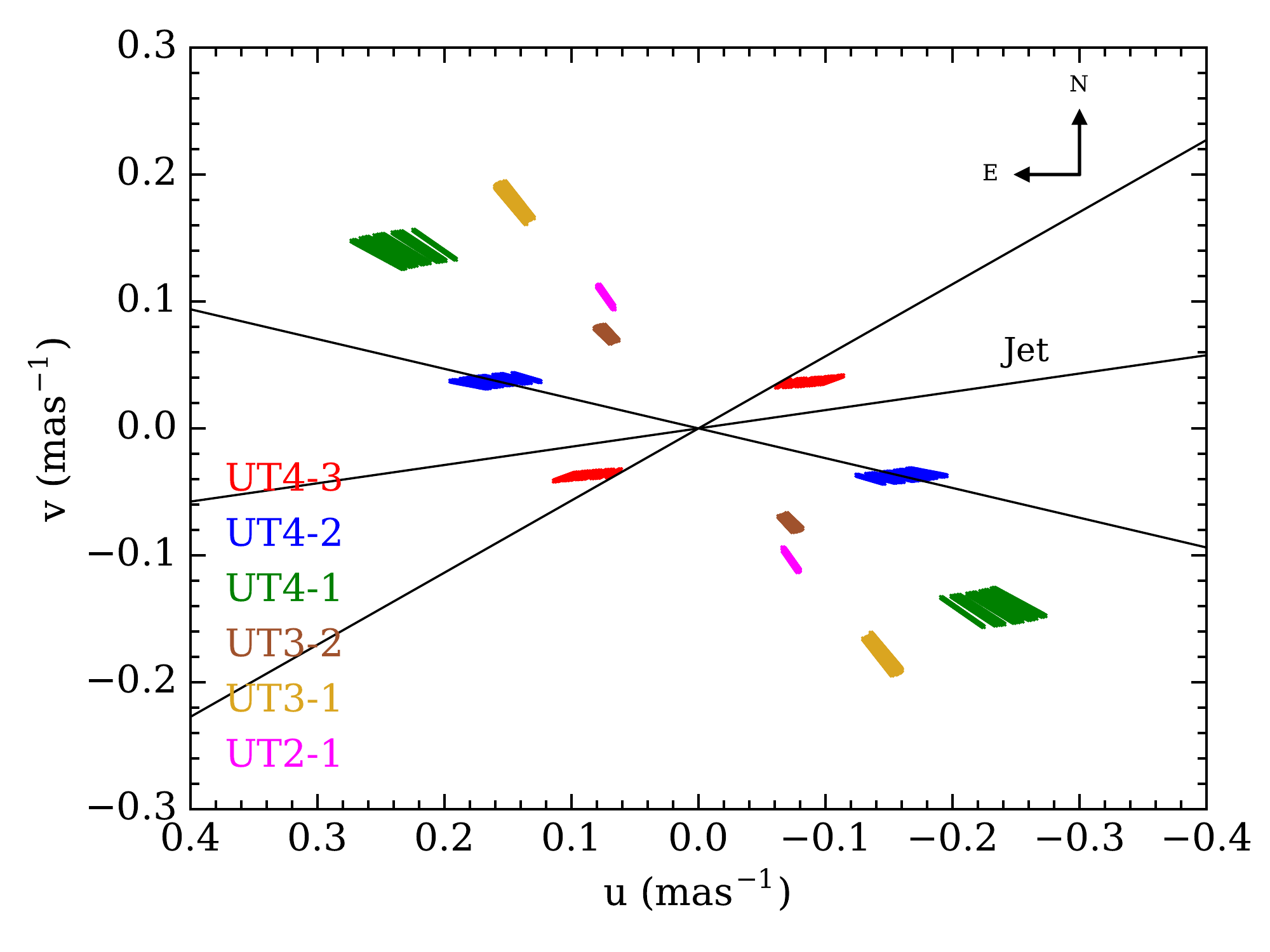} \\ 
\caption{The uv-coverage for Epoch 2 of the GRAVITY 2017 observations. The colors represent the six different baselines, and the coverage in the radial direction corresponds to the different wavelength channels across the K band ($2-2.5 \mu$m). We also show the precessional axis of the jet and its precession cone as seen on sky. The projected baselines are sensitive to both the jet and the orthogonal directions.}
\label{fig:uv_coverage}
\end{figure}

Table \ref{table:obs} summarizes the observations. The precessional phase was $\approx 0.9$, when the disk inclination is close to its minimum value ($i \approx 60^\circ$). The orbital phases varied from $\approx 0.25-0.5$, so that the accretion disk is not eclipsed. Figure \ref{fig:uv_coverage} shows the uv coverage for second observation epoch, with the jet precessional axis and cone as seen in radio observations \citep[e.g.][]{Stirling02}. The uv coverage for the other epochs is similar, but shorter in the third observation. The imaging resolution is $\approx 3$ mas; however, we can resolve structures at sub-mas resolution through spectral differential visibilities. 

\begin{table*}[t]
\centering
\caption{\label{table:obs} Summary of observations.}
\begin{tabular}{ccccccc}
\hline \hline
\shortstack{Date\\Time(UTC)} & \shortstack{Total \\Integration\\Time (min)} & \shortstack{Seeing\\(")} & \shortstack{Coherence Time\\@ 500 nm (ms)} & \shortstack{Calibrator\tablefootmark{a}\\Spectral Type\\Diameter (mas)} & Jet Precessional Phase\tablefootmark{c} & Orbital Phase\tablefootmark{d} \\ [0.3cm]

\shortstack{2017-07-07\\6:25-8:10\\Epoch 1} & $70$ & 0.4-0.6 & 4-6 & \shortstack{HD 183518\\A3V\\$0.157 \pm  0.002$} & 0.895 & 0.252 \\ [0.3cm]

\shortstack{2017-07-09\\6:35-8:10\\Epoch 2} & $60$ & 0.5-0.7 & 8-13 & \shortstack{HD 185440\\A2/3V\tablefootmark{b}\\$0.218 \pm 0.002$} & 0.907 & 0.405 \\ [0.3cm]

\shortstack{2017-07-10\\6:25-6:55\\Epoch 3} & $20$ & 0.4-0.5 & 7-9 & \shortstack{HD 188107\\B9V\\$0.173 \pm 0.002$} & 0.913 & 0.480 \\ [0.3cm]
\hline
\end{tabular}
\tablefoot{\newline
\tablefoottext{a}{Based on \cite{Chelli16}.} \newline
\tablefoottext{b}{This calibrator is probably misclassified as it has strong CO bands in its spectrum.} \newline
\tablefoottext{c}{Based on the kinematic parameters in \cite{Eikenberry01}. Phase zero is when the eastern/western jet is maximally blue/redshifted.} \newline
\tablefoottext{d}{Based on the orbital parameters in \cite{Goranskii98}. Phase zero corresponds to the eclipse center of the accretion disk.} \newline
}
\end{table*}

The data were reduced with the standard GRAVITY pipeline \citep[version 1.0.7,][]{Lapeyrere14}. The interferometric calibrators used are listed in Table \ref{table:obs}. They were also used as tellluric line calibrators. We detected no significant difference in the interferometric quantities between the two polarizations in any of the three nights, either in the continuum FT or differential SC visibilities, and therefore we averaged the two polarizations. The data is also averaged in time for each of the three epochs because we do not see clear variability during each observation. We use the low resolution FT data to study the K band continuum, and the high resolution SC data to study the emission lines through differential visibility amplitudes and phases. Our limited uv coverage resulting from the rather short observations did not allow for model-independent image reconstruction; therefore, we have to rely on model-independent quantities and model fitting. 

\section{The Near-Infrared K band Continuum}
\label{Continuum} 

SS 433 is known to have strong infrared excess \citep[e.g.][]{Allen79,Fuchs06} from extended outflows. At NIR wavelengths, 
we expect a flux contribution from both the accretion disk and donor star ($\ll 0.1 \text{mas}$) and from more extended emission.
In all observations, the continuum closure phases are very small $\lesssim 2^\circ$, pointing to symmetric structures within 
our spatial resolution $\approx 3 \text{ mas}$. 

In Paper I we reported a phenomenological model for the K band continuum consisting of a partially resolved source (FWHM $\lesssim 1 \text{mas}$) 
embedded in a completely resolved background with $\approx 10\%$ of its flux. Here we construct a slightly more involved model in face of the strong evidence for an 
equatorial structure in the 2017 observations. The model consists of two components: an unresolved point source representing the binary (accretion disk + donor star) and a two-dimensional elliptical Gaussian which could represent an extended disk/wind. The model parameters are: 

\begin{enumerate} 

\item the flux ratio $f$ between the Gaussian and the point source components; 

\item the FWHM $\theta_g$ of the Gaussian component along the major axis; 

\item the disk inclination $i$, which gives the aspect ratio of the Gaussian $\cos(i)$; 

\item the position angle (PA) of the Gaussian axis.

\end{enumerate}

The model visibility is therefore: 

\begin{equation}
V (\vect{u}) = \frac{1 + f \times V_{gaussian} (\vect{u})}{1+f} 
\end{equation}

\noindent where $V_{gaussian}$ is the visibility of the elliptical Gaussian and $\vect{u} = \dfrac{\vect{B}}{\lambda}$ with $\vect{B}$ the baseline vector. 

\begin{figure}[tb]
\centering
\includegraphics[width=\columnwidth]{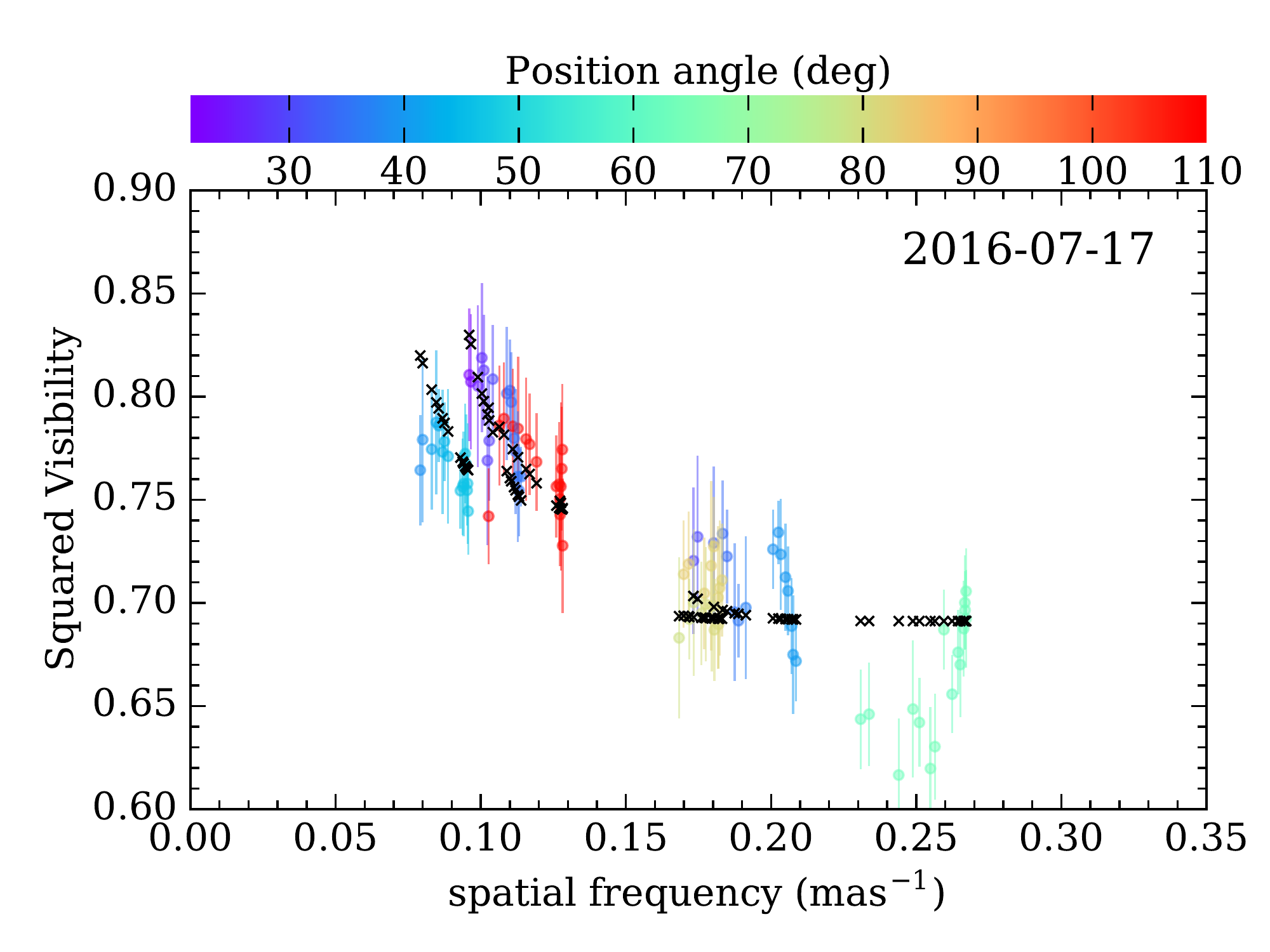} \\ 
\includegraphics[width=\columnwidth]{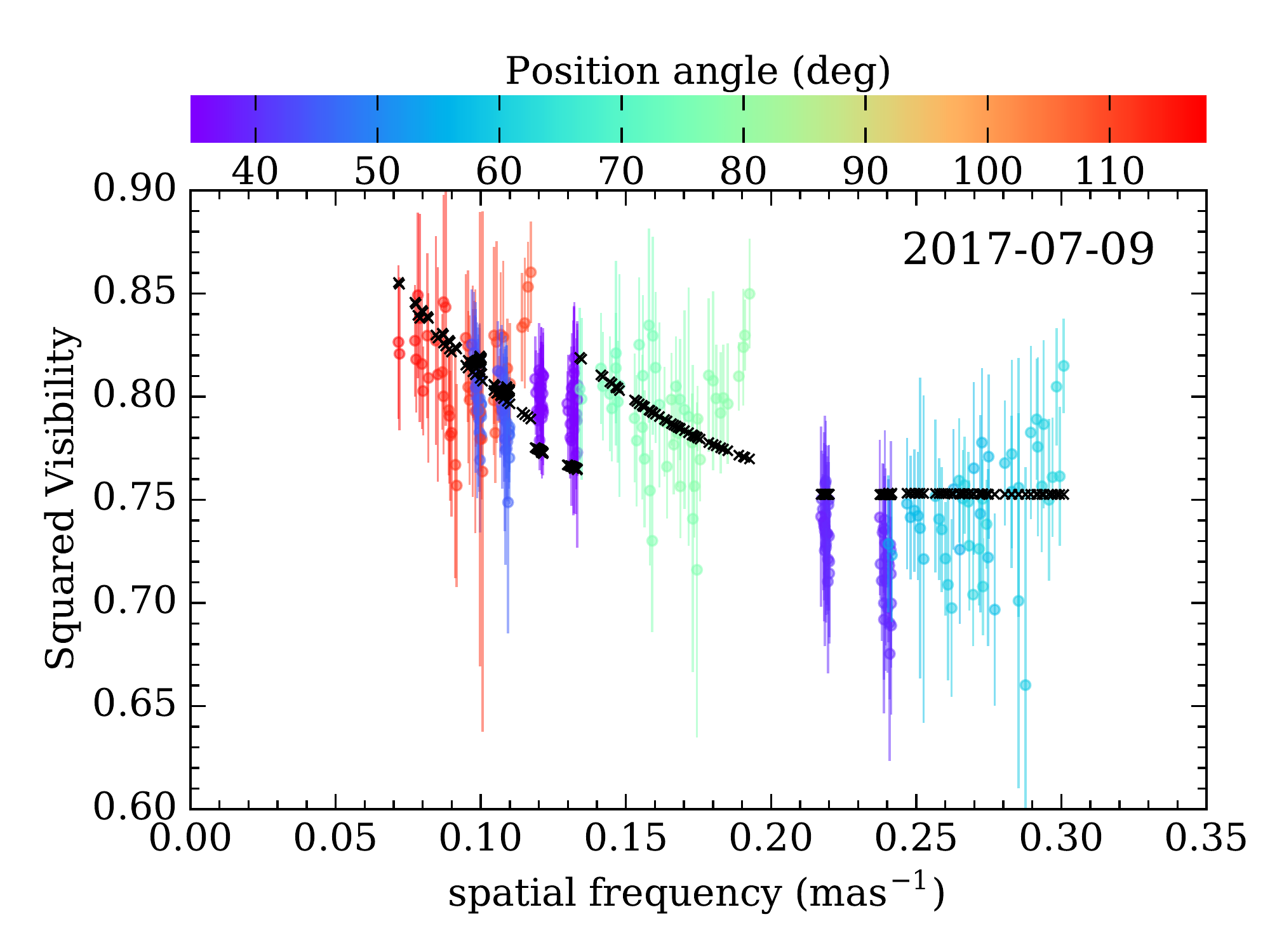} \\ 
\caption{K band continuum squared visibilities of SS 433 from the GRAVITY fringe tracker for the 2016 observation (\textbf{top}) and Epoch 2 of the 2017 observation (\textbf{bottom}). It is modeled by the combination of an unresolved point source representing the binary (accretion disk + donor star) and a two-dimensional elliptical Gaussian which could represent an extended disk/wind. The data is shown in color and the best fit model is shown in black.}
\label{fig:cont}
\end{figure}

Table \ref{table:FT} and Figure \ref{fig:cont} show the results for the model fits to the continuum squared visibilities for the 2016 observation and Epoch 2 of the 2017 observations (the other epochs look similar). Because the measurement errors are dominated by systematics from imperfect calibration of the visibilities (which leads to large $\frac{\chi^2}{\text{dof}}$), we estimate the parameter errors from bootstrapping over the different baselines. We also note that spectral channels with strong emission lines were not used, to avoid the biasing of continuum visibilities by the differential visibilities. 

\begin{table*}[t]
\centering
\caption{\label{table:FT} K Band Continuum Model Fit Results}
\begin{tabular}{cccccc} 
\hline \hline \\ 
Parameter & unit & 2016-07-17 & \shortstack{2017-07-07\\Epoch1} & \shortstack{2017-07-09\\Epoch2} & \shortstack{2017-07-10\\Epoch3} \\ [0.3cm]
f & - & $0.20 \pm 0.01$ & $0.10 \pm 0.01$ & $0.15 \pm 0.01$ & $0.28 \pm 0.01$ \\ [0.3cm]
$\theta_g$ & mas & $7 \pm 1$ & $7 \pm 1$ & $9 \pm 2$ & $7 \pm 1$ \\ [0.3cm] 
$i$ & deg & $90 \pm 27$ & $48 \pm 16$ & $60 \pm 13$ & $90 \pm 14$ \\ [0.3cm]
PA & deg & $22 \pm 58$ & $98 \pm 39$ & $103 \pm 43$ & $12 \pm 63$ \\ [0.3cm]
\hline
$\frac{\chi^2}{dof}$ & & $60$ & $102$ & $42$ & $127$ \\ [0.3cm]
\hline 
\end{tabular}
\end{table*}

The model fits point to an extended structure with a FWHM $\sim 7$ mas containing $10-30 \%$ of the flux of the central point source. The PAs are very not well-constrained, but the inclinations do not favor a symmetric Gaussian. The extended continuum structure could therefore correspond to a disk, with some possible contribution from an extended wind (both of which are seen in the Br$\gamma$ line, see below). The inclination and position angle of the jets are $i_{jet} \approx 90^\circ$, $PA_{jet} \approx 75^\circ$ in the 2016 observation (Paper I) and $i_{jet} \approx 60^\circ$, $PA_{jet} \approx 88^\circ$ for the 2017 observations (see companion paper on the jets, Waisberg et al. sub.). 

\section{The Stationary Br$\gamma$ line} 
\label{Brg}

\subsection*{The K band spectrum} 

The K band spectrum of SS 433 contains both stationary emission lines (Br$\gamma$, He I 2.06 $\mu$m, He I 2.112 $\mu$m and high order (upper levels 19-24) Pfund lines) 
as well as emission lines from the baryonic jets. By far the strongest stationary line is the Br$\gamma$ line, which is the focus of this paper. It is a broad line with FWHM $\sim 1000$ km/s and often shows a double peaked structure. We note that the Br$\gamma$ stationary line in our observations is partially blended with Pa$\alpha$ emission lines from the receding jet. Figure \ref{fig:detail} shows the relevant part of the K band spectrum for Epoch 2, with velocities centered on the Br$\gamma$ line. For the complete K band spectra, we refer to the companion paper on the jets (Waisberg et al. sub.).  

\subsection*{Model-indepedent Results}

\begin{figure*}[]
\centering
\includegraphics[width=1.8\columnwidth]{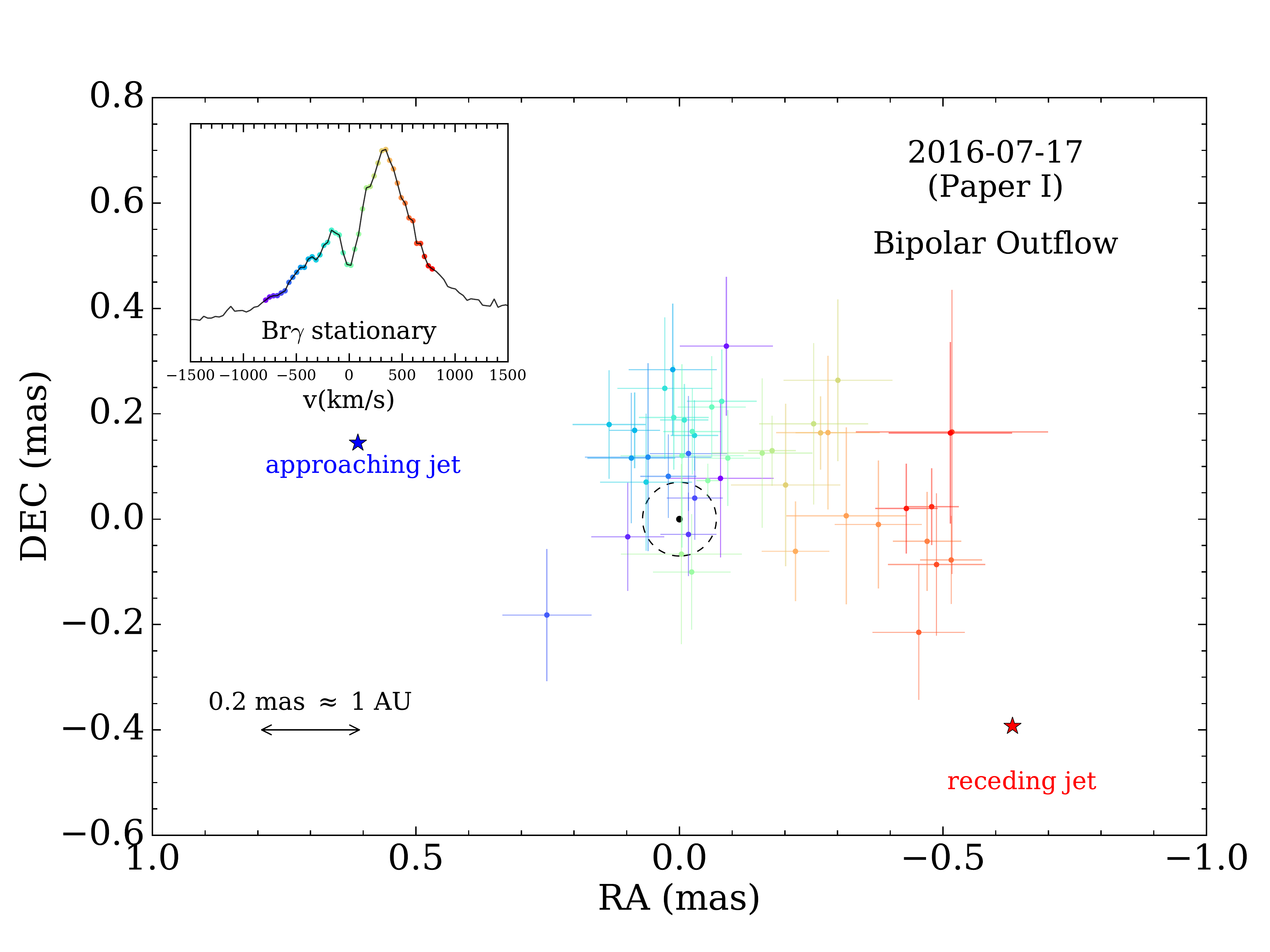} \\ 
\includegraphics[width=1.8\columnwidth]{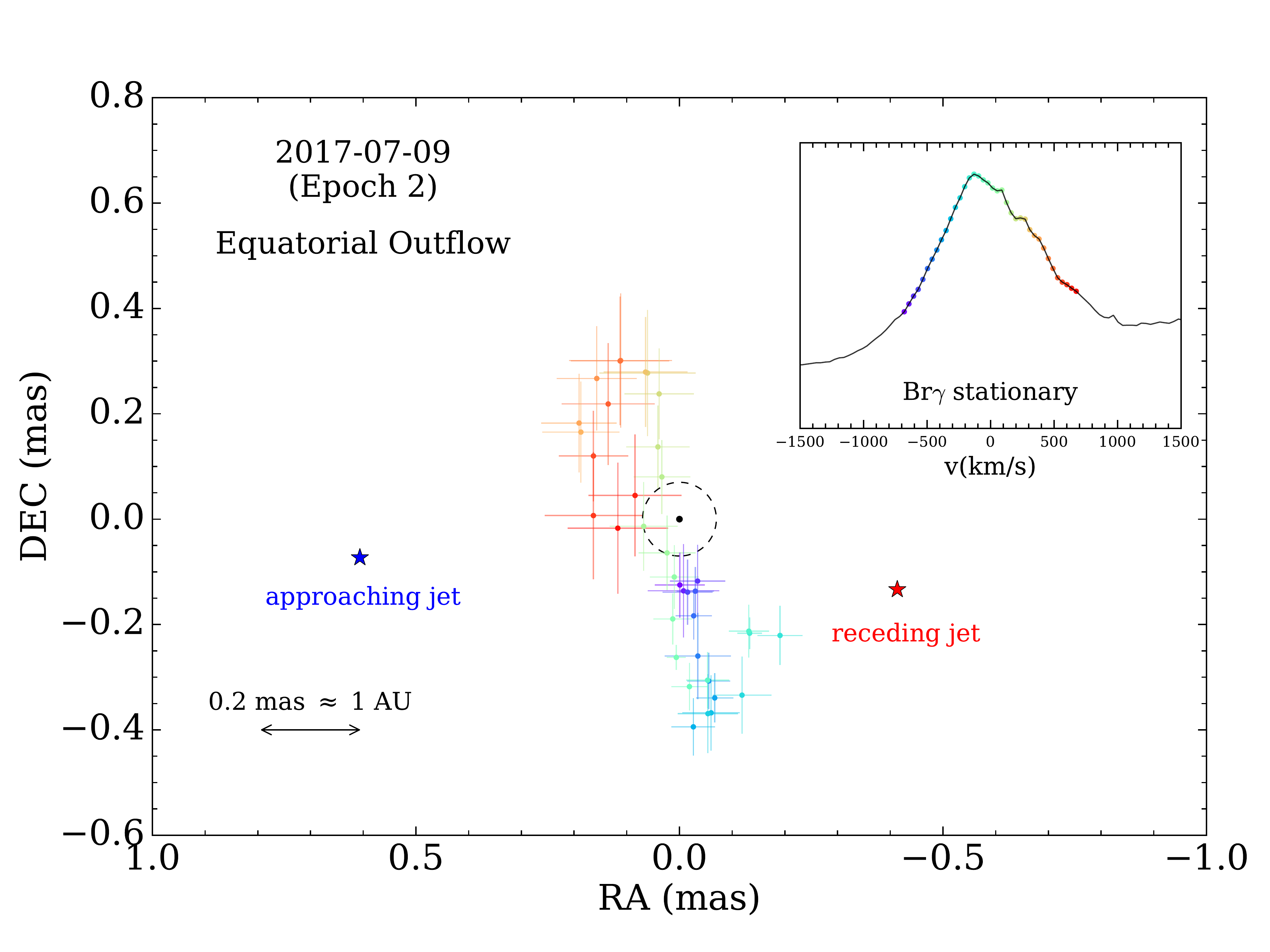} \\ 
\caption{Model-independent centroid shifts across the Br$\gamma$ stationary line for the 2016 observation and Epoch 2 of the 2017 observations. The dashed circle shows the binary size $a \approx 0.07$ mas for a total mass $M=40 M_{\odot}$. The insets show the Br$\gamma$ line spectrum with the color corresponding to the different wavelength/velocity channels.}
\label{fig:centroid} 
\end{figure*}

As mentioned previously, the stationary emission lines in SS 433 have been ascribed to multiple components, including an accretion disk, extended accretion disk wind/outflow and circumbinary ring. Our interferometric data \textit{spatially} resolve the Br$\gamma$ line emission for the first time. The differential phases on most baselines show a remarkable "S-shape", which is a typical signature of a spatial velocity gradient (Figure \ref{fig:detail}). A comparison of the differential phases between the jet lines and the Br$\gamma$ line reveals that in 2017 the latter is perpendicular to the jets, rather than along their direction as was the case in the 2016 observation (Paper I). This can be clearly visualized in a model-independent way by converting the differential phases $\Delta \phi$ to centroid offsets $\Delta \vect{x}$ between the line and the continuum, since in the marginally resolved limit \citep[e.g.][]{Monnier13,Waisberg17}:

\begin{equation}
\Delta \phi = -2 \pi \vect{u} \cdot \Delta \vect{x} \left(\frac{f-1}{f}\right)
\end{equation}

\noindent where $\vect{u} = \dfrac{\vect{B}}{\lambda}$ and $f$ is the line flux in continuum-normalized units. Figure \ref{fig:centroid} shows the centroid of emission across the Br$\gamma$ line for the 2016 observation (Paper I) and Epoch 2 of the 2017 observations, along with the centroid of the jet emission lines. The emission is dominated by a bipolar (jet-like) structure in the 2016 observation, as reported in Paper I (although with a substantial scatter and an apparent offset $\approx 0.2$ mas between the jet PA and the stationary line PA), and by a clear equatorial structure in the 2017 observation. 

Figure 4 shows the interferometric data of Epoch 2 of the 2017 observations on two representative baselines (one close to perpendicular to the jets, the other close to parallel to the jets). We note the following important findings: 

\begin{enumerate}

\item From Figure 3, the higher velocity part of the line is more compact than the lower velocity, which points to a significant rotation component rather than a radially accelerating outflow; 

\item Figure 4 shows that the differential phase peaks, which are much stronger in baselines closer to perpendicular to the jets, have a half-separation of $\sim 250$ km/s, and extend to $\gtrsim 1000$ km/s. The jet inclination in SS 433 is $\gtrsim 60^\circ$, so that any disk-like component is likely to be very close to edge-on, so that this velocity should be close to the deprojected velocity; 

\item From Figure 3, the centroid displacement near the line peak (where the differential phase peaks occur) is $\gtrsim 0.4$ mas. This is a lower limit to the size of the region associated with that intrinsic velocity because, in a disk, projection effects cause inner material to also contribute to that velocity. We also note that, if the emission line has additional components so that the true $f$ is reduced, the centroid displacement necessary to produce the same differential phase will be larger. Therefore, $0.4$ mas is a lower limit to the size of the region where the velocity is $\sim 250$ km/s. 

\end{enumerate} 

This clearly shows that the rotating structure is too extended to be an accretion disk, since $a_{orb} < 0.07$ mas for a total binary mass $M < 40 M_{\odot}$. The phase peaks $\sim 250$ km/s are close in velocity to the Gaussian components that have been associated with a circumbinary ring in previous spectral decompositions \citep{Blundell08,Bowler10}; however, the interferometric data show that these structures are too extended to correspond to the inner edge of a circumbinary ring at $\approx 2 a_{orb}$. We note that the Keplerian velocity $v_{\rm Kep} = \sqrt{\frac{GM}{r}}$ at $0.4$ mas for $M < 40 M_{\odot}$ is $\approx 130$ km/s, so that if in Keplerian rotation the structure would imply a very high enclosed mass $\gtrsim 150 M_{\odot}$. Alternatively, the structure may be an equatorial rotating outflow (see below). 

\subsection*{Models}

We model the equatorial structure emitting in Br$\gamma$ as a geometrically and optically thin disk-like structure, which can be either stationary in Keplerian rotation or expanding. The parameters are as follows:

\begin{enumerate}

\item The outer radius $R_{out}$ (mas), beyond which Br$\gamma$ emission ceases; 

\item The ratio of outer to inner radius $\dfrac{R_{out}}{R_{in}}$, the latter demarcating the radius at which Br$\gamma$ emission begins; 

\item The radial emission profile in Br$\gamma$, parametrized by $I(r) \propto r^{-\alpha}$; 

\item The deprojected rotation velocity at the outer radius, $v_{\phi} (R_{out})$. The rotational velocity is given by 

\begin{align}
v_{\phi} (r) = v_{\phi} (R_{out}) \left(\frac{R_{out}}{r}\right)^\beta
\end{align}

\noindent where $\beta=0.5$ for Keplerian rotation and $\beta=1$ for an expanding outflow from conservation of angular momentum. 

\item The outflow velocity $v_{r}$ for the case $\beta=1$. This is assumed to be constant i.e. the outflow has reached its terminal velocity by the time Br$\gamma$ emission starts. 

\item The inclination $i$ of the disk; 

\item The position angle $PA$ of the disk; 

\item The systematic velocity of the disk, $v_{sys}$, which could include e.g. orbital motion; 

\item The turbulence velocity fraction, given by $\sigma = \frac{v_{turb}}{v_{\phi}}$. This parameter makes the double-peaked profile typical of disks less pronounced. 

\end{enumerate} 

In addition, there is a need for an extended component, which also creates the high velocity $\gtrsim 1000$ km/s wings of the line profile. This is because the differential visibility amplitudes decrease across the line, pointing to a net structure that is more extended than the continuum, whereas the disk alone would cause an increase in visibility amplitude if no other component were present (Figure \ref{fig:detail}). The presence of a broad wind component agrees with previous spectroscopic decompositions of the stationary lines \citep[e.g.][]{Blundell08,Perez09}. We model it as a spherically symmetric component, assumed to produce a Gaussian emission line in the spectrum and a symmetric 2d Gaussian in the image. Since it is spherically symmetric, this component does not induce differential visibility phase shifts. Its model parameters are: 

\begin{enumerate}

\item The strength and FWHM (km/s) of the wind line in the spectrum, $\text{FWHM}_{wind}$; 

\item The size (FWHM) of the wind image (mas), $\theta_{wind}$ 

\end{enumerate} 

\noindent The systemic velocity is assumed to be the same as for the equatorial structure. 

The errors for the science channel are estimated from the scatter in line-free regions. We fit for the spectrum and the differential visibilities simultaneously; however, because the former is sensitive to telluric correction and has very small statistical error bars, we increase the flux error bars by a factor of two. We find that this scaling led to a comparable reduced $\chi^2$ between flux and visibilities in all observations. Moreover, because of the blending with Pa$\alpha$ emission lines from the receding jet, which also produces differential visibility signatures, it is necessary to perform simultaneous fits for the Br$\gamma$ line and the jets. For the model and results for the jets we refer to the companion paper (Waisberg et al., sub.). 

For the velocity-resolved interferometric model for the equatorial disk detailed above, we construct a spatial grid with velocities and fluxes determined by the model parameters, and the visibilities are then computed through a numerical 2D Fourier transform. The total differential visibility at a given spectral channel is then 

\begin{equation}
V_{\rm diff} (\vect{u}) = \frac{V(\vect{u})}{V_c(\vect{u})} = \frac{1 + \sum \limits_{i} \frac{V_i(\vect{u})}{V_c(\vect{u})} f_i}{1 +\sum \limits_{i}  f_i}
\end{equation}

\noindent where $V_c$ is the continuum visibility (taken from the best fit continuum model, Section 3), and $f_i$ and $V_i$ are the flux ratios relative to the continuum and visibilities for each component $i$ (equatorial disk/outflow, extended wind and jets). 

The fits are done through non-linear least squares minimization with the Levenberg-Marquardt method through the \texttt{python} package LMFIT \footnote{https://lmfit.github.io/lmfit-py/}. The quoted errors correspond to the 1-$\sigma$ errors from the least squares fit i.e. the estimated derivatives around the optimal solution (scaled by $\sqrt{\chi^2_{red}}$). We caution, however, that true uncertainties are dominated by (i) degeneracies between the many parameters, which create a complicated multi-dimensional $\chi^2$ map ; (ii) systematic errors from the continuum model; (iii) the assumption of our simple "geometric" models, which cannot capture all the complexities involved. A more realistic assessment of the errors can probably be grasped from the comparison between the results of the three different epochs (barring fast variability). We note, however, that in Epoch 1 there is very severe spectral blending of the different components, so that its results are less robust. 

\subsection*{Model Results} 

\begin{figure*}[]
\centering
\includegraphics[width=2\columnwidth]{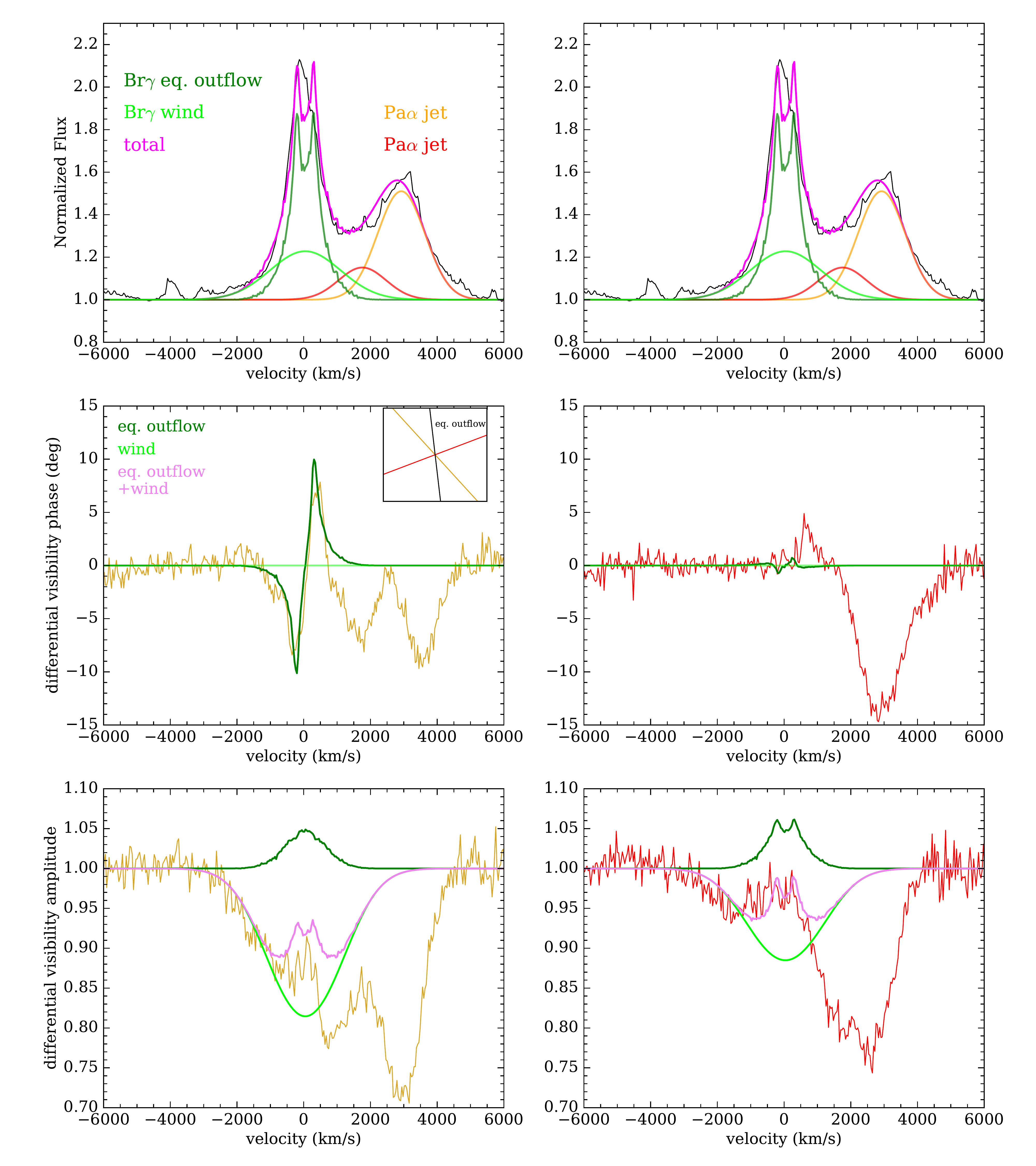}
\caption{This figure illustrates the main features of the data and model. We show two representative baselines (UT3-1, yellow and UT4-3, red -- see Figure \ref{fig:uv_coverage}) for Epoch 2 and the outflow model fit. The \textbf{top} row shows the spectrum centered on the Br$\gamma$ stationary line. The latter is decomposed into an equatorial outflow (dark green) and a spherical wind (lime). The former is responsible for the S-shape signatures in the differential visibility phases (\textbf{middle} row) for baselines which are close to perpendicular to the jet (left), and show almost no signature on baselines more aligned with the jet (right). The inset shows the position angle of the outflow from the fit as well as the baseline directions on the sky plane. The \textbf{bottom} row shows the differential visibility amplitudes. The equatorial outflow alone would lead to an increase in visibility amplitude across the Br$\gamma$ line. The extended wind component can explain both the high velocity wings $\gtrsim 1000$ km/s in the spectrum as well as the net decrease in visibility amplitude across the line. Note that there are two Pa$\alpha$ emission lines from the receding jet which are blended with the Br$\gamma$ stationary line on its red side, and which also create strong visibility signatures. The model fits were done for all the components simultaneously, but here we show only the visibility model for the stationary line for clarity (full model is shown in the Appendix). For the jet model and results, we refer to the companion paper on the jets (Waisberg et al., sub.)}
\label{fig:detail} 
\end{figure*}

Table \ref{table:disk} shows the model fit results for both the disk and outflow models. Figure \ref{fig:detail} shows the data and best fit for the "outflow" model in Epoch 2 for two representative baselines. All the data and best fits for the three epochs for "outflow" model are shown in the Appendix. 

The PA of the equatorial structure is close (although not exactly) perpendicular to the jets (the fit jet PA is $88^\circ$, see the companion paper on the jets, Waisberg et al., sub.), confirming the results from the model-independent analysis. The inclination of the outflow is also close to the jet inclination during the observations ($60^\circ$). Interestingly, the rotation direction of the equatorial outflow is retrograde relative to the jet precession (known from radio and optical observations), in agreement with the slaved disk precession model, according to which the precession is driven from gravitational torques from the compact object on a donor star with a spin axis misaligned with the binary plane \citep{Roberts74,vandenHeuvel80,Whitmire80}. Modeling of the eclipses in the X-ray and optical light curves at different precessional phases had shown evidence of retrograde precession \citep[e.g.][]{Brinkmann89,Leibowitz84}; our results clearly confirm that the jets precess retrograde relatively to the binary. 

The "disk" and "outflow" models look very similar and cannot be distinguished based on the $\chi^2$. However, we disfavor the "disk" model based on the following arguments. The resulting enclosed mass is very high $\sim 400 M_{\odot}$, which follows from the fact that the disk is too extended $R_{out} \approx 1 \text{ mas}$ for its velocity $v_{\phi} (R_{out}) \approx 260 \text{ km/s}$. It would entail that SS 433 harbors an intermediate mass black hole, which is strongly disfavored by all that is known about the object, such as the radial velocity curves and eclipse behavior \citep{Fabrika04}. Even more problematic is the fact that the "disk" model is not self-consistent: for such a high mass, $a_{orb} \sim 0.15 \text{ mas}$, which is larger than the resulting inner radius $\lesssim 0.1 \text{ mas}$; therefore, the disk would have to pass through the binary orbit. 

Instead, we favor the "outflow" model. In this case, the extended equatorial structure we detect would correspond to an outflow moving out at $v_r \sim 240 \text{ km/s}$ but with a very significant rotation component. The inner edge of the outflow at $\sim 0.1 \text{ mas}$ has a rotational velocity $\sim 1500 \text{ km/s}$ and the outer edge at $\approx 0.7 \text{ mas}$ rotates at $ \approx 220 \text{km/s}$. This corresponds to a very high specific angular momentum, which is $\gtrsim 10$ times larger than the specific orbital angular momentum of the compact object $l_X$ for a total mass $M<40 M_{\odot}$, assuming a radial velocity amplitude $K_X = 175$ km/s as derived from the HeII 4686$\AA$ line \citep{Fabrika90} and a binary inclination $i = 78^\circ$ \citep{Eikenberry01}, 

\begin{align}
l_X = \frac{K_X}{\sin i} a_X =  \frac{K_X}{\sin i} \frac{a_{orb}}{1+q} < \frac{K_X}{\sin i} a_{orb} = \\ 178 \text{ km/s} \times 0.07 \text{ mas} \left(\frac{M}{40 M_{\odot}}\right)^{1/3}
\end{align}

\noindent where $q = \frac{M_X}{M_*}$ is the mass ratio. The specific angular momentum of the donor star is even smaller, since $q<1$ based on radial velocities \citep{Hillwig08,Kubota10} or other estimates of the mass ratio \citep{Cherepashchuk18,Bowler18}. The extended outflows therefore seem to require a mechanism for transfer of substantial specific angular momentum to the outflowing material, which requires tidal or magnetic torques. We note that the inferred expansion velocity of the outflow $\approx 240$ km/s corresponds to only $\sim 0.02$ mas/day, so that we do not expect to detect movement within the different epochs within the uncertainties. 

Because the inner radius of the outflow $\sim 0.1 \text{ mas} \sim a_{orb}$, it appears that it is launched from circumbinary material, disfavoring an origin 
from the accretion disk. We also disfavor such an origin based on theoretical arguments: centrifugal outflows from magnetic torques in the accretion disk 
itself require geometrically thin disks ($H \ll R$) threaded by vertical magnetic fields which dominate the pressure, whereas in SS 433 the disk is geometrically 
thick ($H \sim R$) and the extreme mass inflow $\dot{M} = 10^{-4} M_{\odot}/\text{yr}$ creates a large inward ram pressure as well as radiation pressure from  supercritical accretion ($\dot{M} \gg \dot{M}_{Eddington}$). A more exotic possibility would be extraction of angular momentum from a neutron star through a magnetic propeller effect, in which transfer of angular momentum from the spinning neutron star to the flow can happen if the magnetospheric radius $R_m$ is larger than the co-rotation radius $R_{co}$, leading the flow to be centrifugally ejected \citep{Illarionov75}. Although the currently favored model for SS 433 is based on accretion-powered outflows from a massive stellar mass black hole \citep{Fabrika04,Cherepashchuk18,Bowler18}, neutron star models for SS 433 have been considered in the past \citep[e.g.][]{Begelman80,Begelman84}, including the idea that it could be a supercritical propeller \citep{Mineshige91}. In the latter scenario, $R_m$ is smaller than the spherization radius $R_{sp}$ so that a thick disk could still form \citep[$R_{sp}$ is the radius within which the disk becomes geometrically thick from radiation pressure, determined from $L(R>R_{sp}) = L_{Eddington}$;][]{Shakura73}. Our equatorial outflow model would require a launch radius $\gtrsim 4 \times 10^{10} \text{ cm}$ for its speed not to exceed $c$, which is comparable to $R_{sp} \approx 10^{10} \text{ cm} \dfrac{\dot{M}}{10^{-4} M_{\odot}/\text{yr}}$ in SS 433. However, for such a large magnetospheric radius to be inside the light cylinder of a neutron star would require a spin period $P \gtrsim 10 \text{ s}$, deeming a propeller mechanism very unlikely (in addition, it would require very large $\gg 10^{15} \text{ G}$ surface magnetic fields).  

We instead favor that the outflow is driven from a circumbinary disk. There is strong evidence for a such a disk from optical spectroscopy \citep{Blundell08,Bowler10}, which appears to be in Keplerian rotation at $\sim 1.5 a_{orb}$ with a speed of $\sim 250 \text{ km/s}$ \citep{Bowler18}. It is probably fed by excretion through the Lagrangian points behind the compact object, donor star, or both. We note that the specific angular momentum of such disk is a few times higher than available in either binary component, so that its formation must also involve transfer of specific angular momentum. Just the same, there is strong evidence that this circumbinary disk is not stable: we see no evidence for equatorial material in the 2016 GRAVITY observation, where the asymmetric double-peaked Br$\gamma$ line is instead aligned in the jet direction. We suggest that the equatorial structure we detected in optical interferometry traces the inner part of a centrifugally ejected disk, which implies there must be an efficient transfer of specific angular momentum from the binary to the disk, making it super-Keplerian by a factor $\sim 7$, probably through tidal torques \citep[e.g.][]{Chen09}. It is then tempting to associate the enigmatic equatorial outflows sometimes detected in radio images to similar episodes of centrifugal ejection of the circumbinary disk. Interestingly,  \cite{Goranskii17} reports on episodes of disappearance and reappearance of the eclipses and jets in SS 433, which they associate with the formation and ejection of a common envelope, and which could in turn be related to the formation and/or ejection of the circumbinary disk. 

The feeding of the circumbinary disk removes angular momentum from the binary, and \cite{Cherepashchuk18} has recently used the stability of the orbital period reported in \cite{Goranskii11} to constrain the mass ratio $q \gtrsim 0.6$. The ejection of the circumbinary disk we suggest here could also have important implications to the binary evolution. However, because it is most probably a transient structure, more observations are needed to understand its cadence and behavior, and that of the outflows in SS 433 in general. The two optical interferometric observations with GRAVITY so far have revealed extremely variable spatial structure to the line emission. 

Finally, we note that the spherical wind component, with FWHM 2,000-3,000 km/s, surrounds the entire binary with a FWHM size $\sim 5-6 \text{ mas}$ (the fit size of the wind is much smaller for Epoch 1; however, this epoch suffers from severe blending with jet emission lines, so that its parameters are much more degenerate and difficult to constrain). The entire SS 433 system appears to be engulfed in an optically thin line emission envelope. 

\begin{table*}[ht]
\centering
\caption{\label{table:disk} Stationary Br$\gamma$ Model Fit Results}
\begin{tabular}{cccccc} 
\hline \hline \\ 
Parameter & unit & Model & \shortstack{2017-07-07\\Epoch 1} & \shortstack{2017-07-09\\Epoch 2} & \shortstack{2017-07-10\\Epoch 3} \\[0.3cm] 
\hline \\ 
\multicolumn{5}{c}{Equatorial Structure Parameters} \\[0.3cm]

\shortstack{$R_{out}$} & mas & \shortstack{disk\\outflow} & \shortstack{$1.31\pm0.07$\\$0.93\pm0.04$} & \shortstack{$1.00\pm0.06$\\$0.71\pm0.04$} & \shortstack{$1.15\pm0.07$\\$0.65\pm0.03$} \\ [0.3cm] 

\shortstack{$\dfrac{R_{out}}{R_{in}}$} & - & \shortstack{disk\\outflow} & \shortstack{$9.5\pm0.2$\\$10.9\pm0.3$} & \shortstack{$12.7\pm0.2$\\$7.0\pm0.1$} & \shortstack{$12.6\pm0.2$\\$7.3\pm0.1$} \\ [0.3cm] 

\shortstack{$\alpha$} & - & \shortstack{disk\\outflow} & \shortstack{$2.32\pm0.07$\\$0.6\pm0.1$} & \shortstack{$2.38\pm0.05$\\$2.19\pm0.05$} & \shortstack{$2.77\pm0.04$\\$2.45\pm0.04$} \\ [0.3cm] 

\shortstack{$v_{\phi} (R_{out})$} & km/s & \shortstack{disk\\outflow} & \shortstack{$277\pm2$\\$284\pm2$} & \shortstack{$258\pm2$\\$215\pm2$} & \shortstack{$243\pm2$\\$216\pm1$} \\ [0.3cm] 

\shortstack{$v_{r}$} & km/s & \shortstack{disk\\outflow} & \shortstack{-\\$240\pm2$} & \shortstack{-\\$236\pm2$} & \shortstack{-\\$232\pm2$}  \\ [0.3cm] 

\shortstack{$i$} & deg & \shortstack{disk\\outflow} & \shortstack{$72\pm1$\\$69\pm1$} & \shortstack{$72.0\pm0.6$\\$56.7\pm0.4$} & \shortstack{$64.8 \pm0.4$\\$52.2\pm0.2$} \\ [0.3cm]

\shortstack{PA} & deg & \shortstack{disk\\outflow} & \shortstack{$97\pm2$\\$86\pm2$} & \shortstack{$104\pm4$\\$96\pm5$} & \shortstack{$106\pm4$\\$107\pm5$} \\ [0.3cm]

\shortstack{$\sigma$} & - & \shortstack{disk\\outflow} & \shortstack{$0.349\pm0.002$\\$0.395\pm0.001$} & \shortstack{$0.375\pm0.001$\\$0.346\pm0.002$} & \shortstack{$0.356\pm0.002$\\$0.368\pm0.001$} \\ [0.3cm]

\shortstack{$M_{enc}$}\tablefootmark{a} & $M_{\odot}$ & \shortstack{disk\\outflow} &  \shortstack{$621\pm34$\\-} & \shortstack{$416\pm27$\\-} &  \shortstack{$420\pm23$\\-} \\ [0.3cm]

\hline \\ 
\multicolumn{5}{c}{Spherical Wind Parameters} \\[0.3cm]

\shortstack{$\text{FWHM}_{wind}$} & km/s & \shortstack{disk\\outflow} & \shortstack{$2975\pm100$\\$2582\pm112$} & \shortstack{$2810\pm65$\\$2494\pm56$} &  \shortstack{$1952\pm40$\\$1809\pm38$} \\ [0.3cm] 

\shortstack{$\theta_{wind}$} & mas & \shortstack{disk\\outflow} & \shortstack{$0.7\pm0.1$\\$0.7\pm0.1$} & \shortstack{$5.3\pm0.2$\\$5.9\pm0.2$} & \shortstack{$5.9\pm0.2$\\$6.3\pm0.3$} \\ [0.3cm] 

\hline \\
\multicolumn{5}{c}{Common Parameters} \\[0.3cm]

\shortstack{$v_{sys}$} & km/s & \shortstack{disk\\outflow} & \shortstack{$97\pm1$\\$66.1\pm0.5$} & \shortstack{$29.7\pm0.5$\\$47.1\pm0.5$} & \shortstack{$31.9\pm0.3$\\$31.2\pm0.3$} \\ [0.3cm] 

\hline \\ 

$\dfrac{\chi^2}{\text{dof}}$ & - & \shortstack{disk\\outflow} & \shortstack{2.7\\2.7} & \shortstack{1.5\\1.6} & \shortstack{1.0\\1.0} \\ [0.3cm]

\hline

\end{tabular}
\tablefoot{
\tablefoottext{a}{The enclosed mass is computed from $R_{out}$ and $v_{\phi} (R_{out})$ for the case of a Keplerian disk.}
}

\end{table*}

\section{CONCLUSIONS} 
\label{Conclusions}

We have presented a second set of optical interferometry observations of the unique microquasar SS 433 with VLTI/GRAVITY. Here, we focused on the analysis of the near-infrared continuum and the Br$\gamma$ stationary line. 

We summarize our results as follows:

\begin{enumerate}

\item The K band continuum is composed of an unresolved point source (accretion disk+donor star) and an extended FWHM $\sim7 \text{ mas}$ structure. The latter is consistent with being an equatorial disk, but could also have a contribution from an extended spherical wind, both of which are seen in the Br$\gamma$ stationary line; 

\item The model-independent emission centroids across the Br$\gamma$ line clearly point to it being dominated by an equatorial (perpendicular to the jets) structure in the 2017 observations, whereas in the previous GRAVITY observation in 2016 the emission was rather more aligned with the jets, suggestive of a bipolar outflow. 
The rotation direction of the outflow is retrograde relative to the jet precession, in accordance with the slaved disk precession model; 

\item The equatorial structure is very extended and carries a specific angular momentum $\gtrsim 10 \times$ greater than the one in either binary component. If interpreted as a disk in Keplerian rotation, it would imply an implausibly high enclosed mass $\sim 400 M_{\odot}$. We suggest instead that it traces an outflow corresponding to the centrifugal ejection of a circumbinary disk, the existence of which has been inferred from optical spectroscopy. The non-detection of an equatorial structure in the 2016 observation suggests that such a disk can disappear. We suggest that the equatorial outflows typically seen in high-resolution radio images correspond to similar episodes of circumbinary disk ejection. The mechanism driving the specific angular momentum transfer necessary to make the disk super-Keplerian and centrifugally eject it is unclear, but possibly associated with tidal torques from the binary components; 

\item The formation and ejection of the circumbinary disk could have an important effect on the binary evolution of SS 433 depending on their cadence. Future optical and radio interferometric observations capable of spatially resolving the outflows are needed to further study them; 

\item In addition to the equatorial structure, the data also suggest a line component from a symmetric and extended spherical wind $\sim 6 \text{ mas}$ responsible for the high-velocity wings $\gtrsim 1000 \text{ km/s}$ of the line. The binary appears therefore to be engulfed in an optically thin and extended emission line envelope. 

\end{enumerate}

\begin{acknowledgements}
We thank the GRAVITY Co-Is, the GRAVITY Consortium and ESO for developing and operating the GRAVITY instrument. In particular, I.W. and J.D. thank the MPE GRAVITY team, in particular F. Eisenhauer, R. Genzel, S.Gillessen, T. Ott, O. Pfhul and E. Sturm. We also thank the GRAVITY team members (W. Brandner, F. Einsenhauer, S. Hippler, M. Horrobin, T. Ott, T. Paumard, O. Pfhul, O. Straub, E. Wieprecht) and ESO staff who were on the mountain during the observations. We also thank P. Kervella for comments on the paper. POP  acknowledges financial support from the CNRS High Energy National Program (PNHE). POP and GD  acknowledge financial support from the CNES. This research has made use of the Jean-Marie Mariotti Center \texttt{SearchCal} service \footnote{Available at http://www.jmmc.fr/searchcal} co-developped by LAGRANGE and IPAG, CDS Astronomical Databases SIMBAD and VIZIER \footnote{Available at http://cdsweb.u-strasbg.fr/}, NASA's Astrophysics Data System Bibliographic Services, NumPy \citep{van2011numpy} and matplotlib, a Python library for publication quality graphics \citep{Hunter2007}. 
\end{acknowledgements}

\bibliographystyle{aa}
\bibliography{mybib.bib}

\begin{thebibliography}{54}
\expandafter\ifx\csname natexlab\endcsname\relax\def\natexlab#1{#1}\fi

\bibitem[{{Allen}(1979)}]{Allen79}
{Allen}, D.~A. 1979, \nat, 281, 284

\bibitem[{{Begelman} \& {Rees}(1984)}]{Begelman84}
{Begelman}, M.~C. \& {Rees}, M.~J. 1984, \mnras, 206, 209

\bibitem[{{Begelman} {et~al.}(1980){Begelman}, {Sarazin}, {Hatchett}, {McKee},
  \& {Arons}}]{Begelman80}
{Begelman}, M.~C., {Sarazin}, C.~L., {Hatchett}, S.~P., {McKee}, C.~F., \&
  {Arons}, J. 1980, \apj, 238, 722

\bibitem[{{Blundell} \& {Bowler}(2004)}]{Blundell04}
{Blundell}, K.~M. \& {Bowler}, M.~G. 2004, \apjl, 616, L159

\bibitem[{{Blundell} {et~al.}(2008){Blundell}, {Bowler}, \&
  {Schmidtobreick}}]{Blundell08}
{Blundell}, K.~M., {Bowler}, M.~G., \& {Schmidtobreick}, L. 2008, \apjl, 678,
  L47

\bibitem[{{Blundell} {et~al.}(2001){Blundell}, {Mioduszewski}, {Muxlow},
  {Podsiadlowski}, \& {Rupen}}]{Blundell01}
{Blundell}, K.~M., {Mioduszewski}, A.~J., {Muxlow}, T.~W.~B., {Podsiadlowski},
  P., \& {Rupen}, M.~P. 2001, \apjl, 562, L79

\bibitem[{{Bowler}(2010)}]{Bowler10}
{Bowler}, M.~G. 2010, \aap, 521, A81

\bibitem[{{Bowler}(2018)}]{Bowler18}
{Bowler}, M.~G. 2018, \aap, 619, L4

\bibitem[{{Brinkmann} {et~al.}(1989){Brinkmann}, {Kawai}, \&
  {Matsuoka}}]{Brinkmann89}
{Brinkmann}, W., {Kawai}, N., \& {Matsuoka}, M. 1989, \aap, 218, L13

\bibitem[{{Chelli} {et~al.}(2016){Chelli}, {Duvert}, {Bourg{\`e}s}, {Mella},
  {Lafrasse}, {Bonneau}, \& {Chesneau}}]{Chelli16}
{Chelli}, A., {Duvert}, G., {Bourg{\`e}s}, L., {et~al.} 2016, \aap, 589, A112

\bibitem[{Chen \& Zeng(2009)}]{Chen09}
Chen, W. \& Zeng, Q. 2009, Chinese Science Bulletin, 54, 711

\bibitem[{{Cherepashchuk} {et~al.}(2018){Cherepashchuk}, {Postnov}, \&
  {Belinski}}]{Cherepashchuk18}
{Cherepashchuk}, A.~M., {Postnov}, K.~A., \& {Belinski}, A.~A. 2018, \mnras,
  479, 4844

\bibitem[{{Clark} \& {Murdin}(1978)}]{Clark78}
{Clark}, D.~H. \& {Murdin}, P. 1978, \nat, 276, 44

\bibitem[{{Doolin} \& {Blundell}(2009)}]{Doolin09}
{Doolin}, S. \& {Blundell}, K.~M. 2009, \apjl, 698, L23

\bibitem[{{Eikenberry} {et~al.}(2001){Eikenberry}, {Cameron}, {Fierce}, {Kull},
  {Dror}, {Houck}, \& {Margon}}]{Eikenberry01}
{Eikenberry}, S.~S., {Cameron}, P.~B., {Fierce}, B.~W., {et~al.} 2001, \apj,
  561, 1027

\bibitem[{{Fabian} \& {Rees}(1979)}]{Fabian79}
{Fabian}, A.~C. \& {Rees}, M.~J. 1979, \mnras, 187, 13P

\bibitem[{{Fabrika}(2004)}]{Fabrika04}
{Fabrika}, S. 2004, Astrophysics and Space Physics Reviews, 12, 1

\bibitem[{{Fabrika}(1993)}]{Fabrika93}
{Fabrika}, S.~N. 1993, \mnras, 261, 241

\bibitem[{{Fabrika} \& {Bychkova}(1990)}]{Fabrika90}
{Fabrika}, S.~N. \& {Bychkova}, L.~V. 1990, \aap, 240, L5

\bibitem[{{Filippenko} {et~al.}(1988){Filippenko}, {Romani}, {Sargent}, \&
  {Blandford}}]{Fillipenko88}
{Filippenko}, A.~V., {Romani}, R.~W., {Sargent}, W.~L.~W., \& {Blandford},
  R.~D. 1988, \aj, 96, 242

\bibitem[{{Fuchs} {et~al.}(2006){Fuchs}, {Koch Miramond}, \&
  {{\'A}brah{\'a}m}}]{Fuchs06}
{Fuchs}, Y., {Koch Miramond}, L., \& {{\'A}brah{\'a}m}, P. 2006, \aap, 445,
  1041

\bibitem[{{Goranskii} {et~al.}(1998){Goranskii}, {Esipov}, \&
  {Cherepashchuk}}]{Goranskii98}
{Goranskii}, V.~P., {Esipov}, V.~F., \& {Cherepashchuk}, A.~M. 1998, Astronomy
  Reports, 42, 209

\bibitem[{{Goranskij}(2011)}]{Goranskii11}
{Goranskij}, V. 2011, Peremennye Zvezdy, 31 [\eprint[arXiv]{1110.5304}]

\bibitem[{{Goranskij}(2017)}]{Goranskii17}
{Goranskij}, V.~P. 2017, in Astronomical Society of the Pacific Conference
  Series, Vol. 510, Stars: From Collapse to Collapse, ed. Y.~Y. {Balega}, D.~O.
  {Kudryavtsev}, I.~I. {Romanyuk}, \& I.~A. {Yakunin}, 466

\bibitem[{{Gravity Collaboration} {et~al.}(2017{\natexlab{a}}){Gravity
  Collaboration}, {Abuter}, {Accardo}, {Amorim}, {Anugu}, {{\'A}vila},
  {Azouaoui}, {Benisty}, {Berger}, {Blind}, {Bonnet}, {Bourget}, {Brandner},
  {Brast}, {Buron}, {Burtscher}, {Cassaing}, {Chapron}, {Choquet},
  {Cl{\'e}net}, {Collin}, {Coud{\'e} Du Foresto}, {de Wit}, {de Zeeuw}, {Deen},
  {Delplancke-Str{\"o}bele}, {Dembet}, {Derie}, {Dexter}, {Duvert}, {Ebert},
  {Eckart}, {Eisenhauer}, {Esselborn}, {F{\'e}dou}, {Finger}, {Garcia}, {Garcia
  Dabo}, {Garcia Lopez}, {Gendron}, {Genzel}, {Gillessen}, {Gonte}, {Gordo},
  {Grould}, {Gr{\"o}zinger}, {Guieu}, {Haguenauer}, {Hans}, {Haubois}, {Haug},
  {Haussmann}, {Henning}, {Hippler}, {Horrobin}, {Huber}, {Hubert}, {Hubin},
  {Hummel}, {Jakob}, {Janssen}, {Jochum}, {Jocou}, {Kaufer}, {Kellner},
  {Kendrew}, {Kern}, {Kervella}, {Kiekebusch}, {Klein}, {Kok}, {Kolb}, {Kulas},
  {Lacour}, {Lapeyr{\`e}re}, {Lazareff}, {Le Bouquin}, {L{\`e}na}, {Lenzen},
  {L{\'e}v{\^e}que}, {Lippa}, {Magnard}, {Mehrgan}, {Mellein}, {M{\'e}rand},
  {Moreno-Ventas}, {Moulin}, {M{\"u}ller}, {M{\"u}ller}, {Neumann}, {Oberti},
  {Ott}, {Pallanca}, {Panduro}, {Pasquini}, {Paumard}, {Percheron}, {Perraut},
  {Perrin}, {Pfl{\"u}ger}, {Pfuhl}, {Phan Duc}, {Plewa}, {Popovic}, {Rabien},
  {Ram{\'{\i}}rez}, {Ramos}, {Rau}, {Riquelme}, {Rohloff}, {Rousset},
  {Sanchez-Bermudez}, {Scheithauer}, {Sch{\"o}ller}, {Schuhler}, {Spyromilio},
  {Straubmeier}, {Sturm}, {Suarez}, {Tristram}, {Ventura}, {Vincent},
  {Waisberg}, {Wank}, {Weber}, {Wieprecht}, {Wiest}, {Wiezorrek}, {Wittkowski},
  {Woillez}, {Wolff}, {Yazici}, {Ziegler}, \& {Zins}}]{GRAVITY17}
{Gravity Collaboration}, {Abuter}, R., {Accardo}, M., {et~al.}
  2017{\natexlab{a}}, \aap, 602, A94

\bibitem[{{Gravity Collaboration} {et~al.}(2017{\natexlab{b}}){Gravity
  Collaboration}, {Petrucci}, {Waisberg}, {Le Bouquin}, {Dexter}, {Dubus},
  {Perraut}, {Kervella}, {Abuter}, {Amorim}, {Anugu}, {Berger}, {Blind},
  {Bonnet}, {Brandner}, {Buron}, {Choquet}, {Cl{\'e}net}, {de Wit}, {Deen},
  {Eckart}, {Eisenhauer}, {Finger}, {Garcia}, {Garcia Lopez}, {Gendron},
  {Genzel}, {Gillessen}, {Gonte}, {Haubois}, {Haug}, {Haussmann}, {Henning},
  {Hippler}, {Horrobin}, {Hubert}, {Jochum}, {Jocou}, {Kok}, {Kolb}, {Kulas},
  {Lacour}, {Lazareff}, {L{\`e}na}, {Lippa}, {M{\'e}rand}, {M{\"u}ller}, {Ott},
  {Panduro}, {Paumard}, {Perrin}, {Pfuhl}, {Ramos}, {Rau}, {Rohloff},
  {Rousset}, {Sanchez-Bermudez}, {Scheithauer}, {Sch{\"o}ller}, {Straubmeier},
  {Sturm}, {Vincent}, {Wank}, {Wieprecht}, {Wiest}, {Wiezorrek}, {Wittkowski},
  {Woillez}, {Yazici}, \& {Zins}}]{GRAVITYSS43317}
{Gravity Collaboration}, {Petrucci}, P.-O., {Waisberg}, I., {et~al.}
  2017{\natexlab{b}}, \aap, 602, L11

\bibitem[{{Hillwig} \& {Gies}(2008)}]{Hillwig08}
{Hillwig}, T.~C. \& {Gies}, D.~R. 2008, \apjl, 676, L37

\bibitem[{Hunter(2007)}]{Hunter2007}
Hunter, J.~D. 2007, Computing In Science \& Engineering, 9, 90

\bibitem[{{Illarionov} \& {Sunyaev}(1975)}]{Illarionov75}
{Illarionov}, A.~F. \& {Sunyaev}, R.~A. 1975, \aap, 39, 185

\bibitem[{{Kubota} {et~al.}(2010){Kubota}, {Ueda}, {Fabrika}, {Medvedev},
  {Barsukova}, {Sholukhova}, \& {Goranskij}}]{Kubota10}
{Kubota}, K., {Ueda}, Y., {Fabrika}, S., {et~al.} 2010, \apj, 709, 1374

\bibitem[{{Lapeyrere} {et~al.}(2014){Lapeyrere}, {Kervella}, {Lacour},
  {Azouaoui}, {Garcia-Dabo}, {Perrin}, {Eisenhauer}, {Perraut}, {Straubmeier},
  {Amorim}, \& {Brandner}}]{Lapeyrere14}
{Lapeyrere}, V., {Kervella}, P., {Lacour}, S., {et~al.} 2014, in \procspie,
  Vol. 9146, Optical and Infrared Interferometry IV, 91462D

\bibitem[{{Leibowitz}(1984)}]{Leibowitz84}
{Leibowitz}, E.~M. 1984, \mnras, 210, 279

\bibitem[{{Luri} {et~al.}(2018){Luri}, {Brown}, {Sarro}, {Arenou},
  {Bailer-Jones}, {Castro-Ginard}, {de Bruijne}, {Prusti}, {Babusiaux}, \&
  {Delgado}}]{Luri18}
{Luri}, X., {Brown}, A.~G.~A., {Sarro}, L.~M., {et~al.} 2018, \aap, 616, A9

\bibitem[{{Margon}(1984)}]{Margon84}
{Margon}, B. 1984, \araa, 22, 507

\bibitem[{{Margon} {et~al.}(1979){Margon}, {Ford}, {Grandi}, \&
  {Stone}}]{Margon79}
{Margon}, B., {Ford}, H.~C., {Grandi}, S.~A., \& {Stone}, R.~P.~S. 1979, \apjl,
  233, L63

\bibitem[{{Marshall} {et~al.}(2013){Marshall}, {Canizares}, {Hillwig},
  {Mioduszewski}, {Rupen}, {Schulz}, {Nowak}, \& {Heinz}}]{Marshall13}
{Marshall}, H.~L., {Canizares}, C.~R., {Hillwig}, T., {et~al.} 2013, \apj, 775,
  75

\bibitem[{{Mineshige} {et~al.}(1991){Mineshige}, {Rees}, \&
  {Fabian}}]{Mineshige91}
{Mineshige}, S., {Rees}, M.~J., \& {Fabian}, A.~C. 1991, \mnras, 251, 555

\bibitem[{{Mioduszewski} {et~al.}(2004){Mioduszewski}, {Rupen}, {Walker},
  {Schillemat}, \& {Taylor}}]{Mioduszewski04}
{Mioduszewski}, A.~J., {Rupen}, M.~P., {Walker}, R.~C., {Schillemat}, K.~M., \&
  {Taylor}, G.~B. 2004, in Bulletin of the American Astronomical Society,
  Vol.~36, AAS/High Energy Astrophysics Division \#8, 967

\bibitem[{{Monnier} \& {Allen}(2013)}]{Monnier13}
{Monnier}, J.~D. \& {Allen}, R.~J. 2013, {Radio and Optical Interferometry:
  Basic Observing Techniques and Data Analysis}, ed. T.~D. {Oswalt} \& H.~E.
  {Bond}, 325

\bibitem[{{Paragi} {et~al.}(2002){Paragi}, {Fejes}, {Vermeulen}, {Schilizzi},
  {Spencer}, \& {Stirling}}]{Paragi02}
{Paragi}, Z., {Fejes}, I., {Vermeulen}, R.~C., {et~al.} 2002, in Proceedings of
  the 6th EVN Symposium, ed. E.~{Ros}, R.~W. {Porcas}, A.~P. {Lobanov}, \&
  J.~A. {Zensus}, 263

\bibitem[{{Paragi} {et~al.}(1999){Paragi}, {Vermeulen}, {Fejes}, {Schilizzi},
  {Spencer}, \& {Stirling}}]{Paragi99}
{Paragi}, Z., {Vermeulen}, R.~C., {Fejes}, I., {et~al.} 1999, \aap, 348, 910

\bibitem[{{Perez M.} \& {Blundell}(2009)}]{Perez09}
{Perez M.}, S. \& {Blundell}, K.~M. 2009, \mnras, 397, 849

\bibitem[{{Roberts}(1974)}]{Roberts74}
{Roberts}, W.~J. 1974, \apj, 187, 575

\bibitem[{{Robinson} {et~al.}(2017){Robinson}, {Froning}, {Jaffe}, {Kaplan},
  {Kim}, {Mace}, {Sokal}, \& {Lee}}]{Robinson17}
{Robinson}, E.~L., {Froning}, C.~S., {Jaffe}, D.~T., {et~al.} 2017, \apj, 841,
  79

\bibitem[{{Shakura} \& {Sunyaev}(1973)}]{Shakura73}
{Shakura}, N.~I. \& {Sunyaev}, R.~A. 1973, \aap, 24, 337

\bibitem[{{Shklovskii}(1981)}]{Shklovskii81}
{Shklovskii}, I.~S. 1981, \sovast, 25, 315

\bibitem[{{Stephenson} \& {Sanduleak}(1977)}]{Stephenson77}
{Stephenson}, C.~B. \& {Sanduleak}, N. 1977, \apjs, 33, 459

\bibitem[{{Stirling} {et~al.}(2002){Stirling}, {Jowett}, {Spencer}, {Paragi},
  {Ogley}, \& {Cawthorne}}]{Stirling02}
{Stirling}, A.~M., {Jowett}, F.~H., {Spencer}, R.~E., {et~al.} 2002, \mnras,
  337, 657

\bibitem[{{van den Heuvel} {et~al.}(1980){van den Heuvel}, {Ostriker}, \&
  {Petterson}}]{vandenHeuvel80}
{van den Heuvel}, E.~P.~J., {Ostriker}, J.~P., \& {Petterson}, J.~A. 1980,
  \aap, 81, L7

\bibitem[{Van Der~Walt {et~al.}(2011)Van Der~Walt, Colbert, \&
  Varoquaux}]{van2011numpy}
Van Der~Walt, S., Colbert, S.~C., \& Varoquaux, G. 2011, Computing in Science
  \& Engineering, 13, 22

\bibitem[{{Vermeulen} {et~al.}(1993){Vermeulen}, {Schilizzi}, {Spencer},
  {Romney}, \& {Fejes}}]{Vermeulen93}
{Vermeulen}, R.~C., {Schilizzi}, R.~T., {Spencer}, R.~E., {Romney}, J.~D., \&
  {Fejes}, I. 1993, \aap, 270, 177

\bibitem[{{Waisberg} {et~al.}(2017){Waisberg}, {Dexter}, {Pfuhl}, {Abuter},
  {Amorim}, {Anugu}, {Berger}, {Blind}, {Bonnet}, {Brandner}, {Buron},
  {Cl{\'e}net}, {de Wit}, {Deen}, {Delplancke-Str{\"o}bele}, {Dembet},
  {Duvert}, {Eckart}, {Eisenhauer}, {F{\'e}dou}, {Finger}, {Garcia}, {Garcia
  Lopez}, {Gendron}, {Genzel}, {Gillessen}, {Haubois}, {Haug}, {Haussmann},
  {Henning}, {Hippler}, {Horrobin}, {Hubert}, {Jochum}, {Jocou}, {Kervella},
  {Kok}, {Kulas}, {Lacour}, {Lapeyr{\`e}re}, {Le Bouquin}, {L{\'e}na}, {Lippa},
  {M{\'e}rand}, {M{\"u}ller}, {Ott}, {Pallanca}, {Panduro}, {Paumard},
  {Perraut}, {Perrin}, {Rabien}, {Ram{\'{\i}}rez}, {Ramos}, {Rau}, {Rohloff},
  {Rousset}, {Sanchez-Bermudez}, {Scheithauer}, {Sch{\"o}ller}, {Straubmeier},
  {Sturm}, {Vincent}, {Wank}, {Wieprecht}, {Wiest}, {Wiezorrek}, {Wittkowski},
  {Woillez}, {Yazici}, \& {GRAVITY Collaboration}}]{Waisberg17}
{Waisberg}, I., {Dexter}, J., {Pfuhl}, O., {et~al.} 2017, \apj, 844, 72

\bibitem[{{Whitmire} \& {Matese}(1980)}]{Whitmire80}
{Whitmire}, D.~P. \& {Matese}, J.~J. 1980, \mnras, 193, 707

\bibitem[{{Zwitter} {et~al.}(1991){Zwitter}, {Calvani}, \&
  {D'Odorico}}]{Zwitter91}
{Zwitter}, T., {Calvani}, M., \& {D'Odorico}, S. 1991, \aap, 251, 92

\end{thebibliography}

\begin{appendix}
\onecolumn 

\section{Full Data and Model Fits} 

Here we show the data (spectrum, differential visibility phases and amplitudes) and best fits for the "outflow" model for the three epochs of the 2017 observations and for all baselines. The solid lines show the models without the blended jet emission lines, whereas the dashed lines show the full combined model (as the fits are done). For the jet models and results, we refer to the companion paper on the jets, Waisberg et al., sub. The projected length and position angle of each baseline is indicated.

\begin{figure*}[tb]
\label{} 
\centering
\includegraphics[width=0.9\columnwidth]{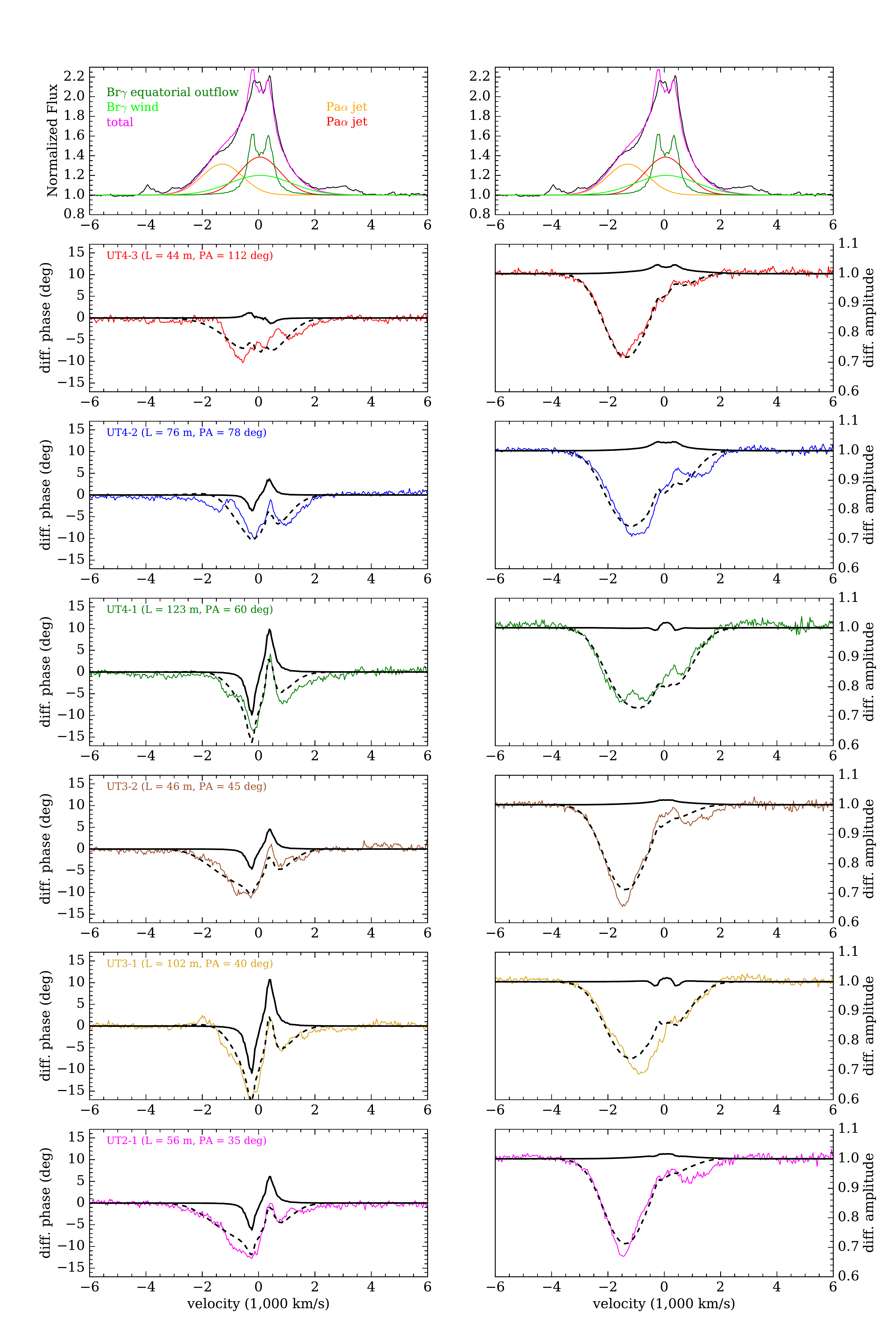}
\caption{Data and best fit "outflow" model for Epoch 1.}
\end{figure*}

\begin{figure*}[tb]
\label{} 
\centering
\includegraphics[width=0.9\columnwidth]{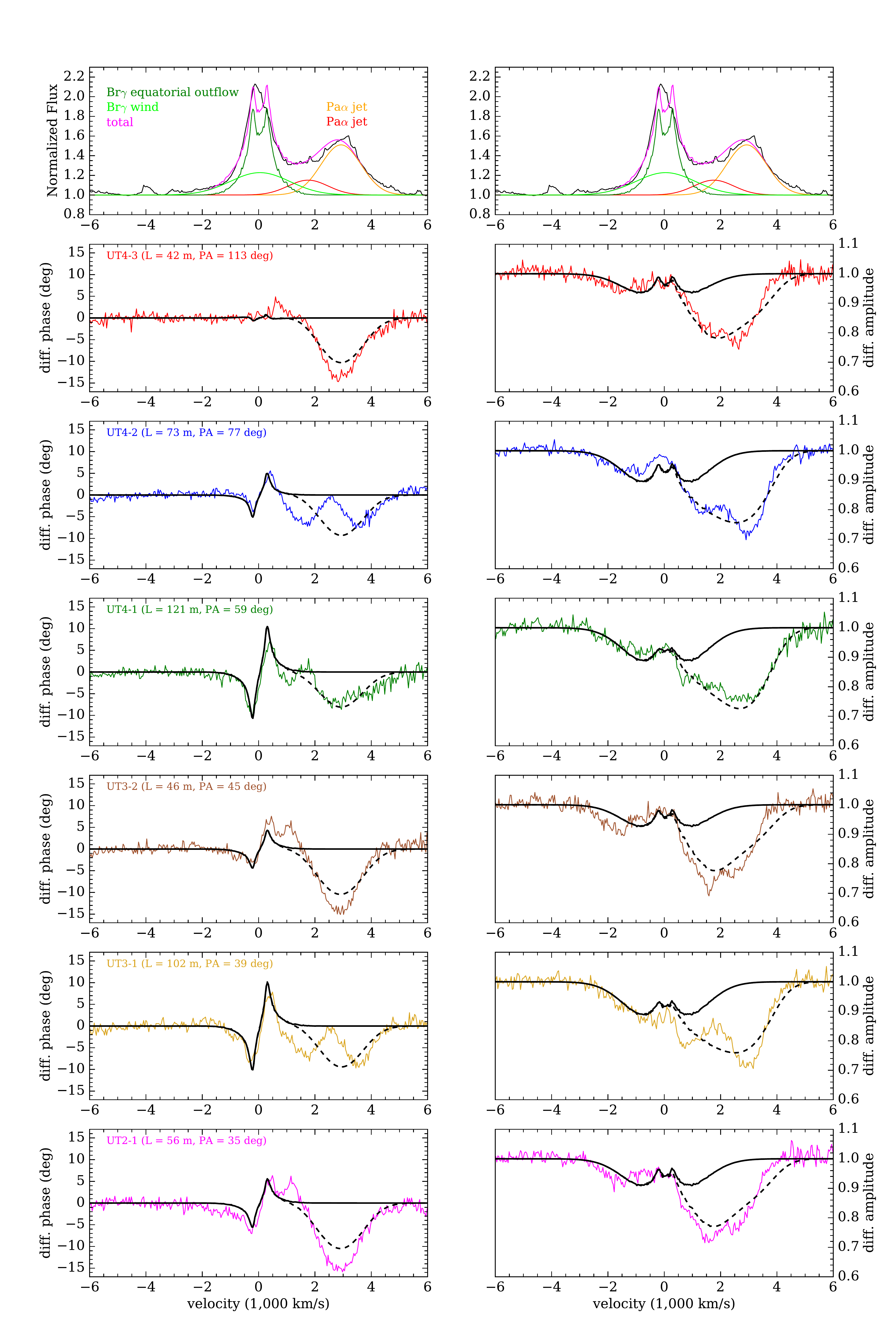}
\caption{Data and best fit "outflow" model for Epoch 2.}
\end{figure*}

\begin{figure*}[tb]
\label{} 
\centering
\includegraphics[width=0.9\columnwidth]{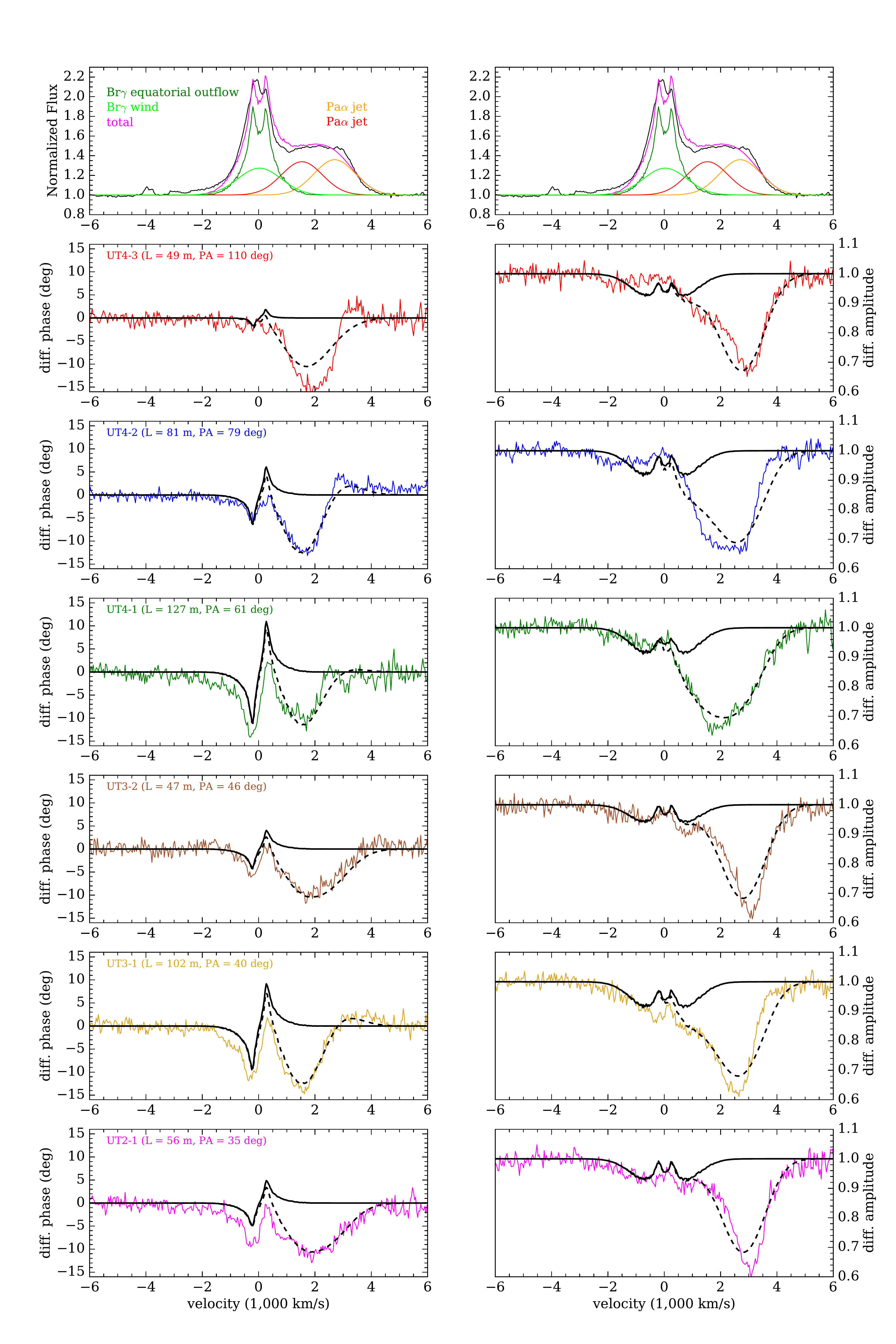}
\caption{Data and best fit "outflow" model for Epoch 3.}
\end{figure*}

\end{appendix}

\end{document}